# A versatile multilayer liquid-liquid encapsulation technique

Utsab Banerjee[1], Sirshendu Misra[1], Sushanta K. Mitra[*]

[1]Equal contribution

Micro & Nano-scale Transport Laboratory, Waterloo Institute for Nanotechnology, Department of Mechanical and Mechatronics Engineering, University of Waterloo, 200 University Avenue West, Waterloo, Ontario N2L 3G1, Canada

[*]Corresponding author: skmitra@uwaterloo.ca

## Abstract

*Hypothesis:* Generating multi-layer cargo using conventional methods is challenging. We hypothesize that incorporating a Y-junction compound droplet generator to encase a target core inside a second liquid can circumvent the kinetic energy dependence of the impact-driven liquid-liquid encapsulation technique, enabling minimally restrictive multi-layer encapsulation.

*Experiments:* Stable wrapping is obtained by impinging a compound droplet (generated using Y-junction) on an interfacial layer of another shell-forming liquid floating on a host liquid bath, leading to double-layered encapsulation. The underlying dynamics of the liquid-liquid interfaces are captured using high-speed imaging. To demonstrate the versatility of the technique, we used various liquids as interfacial layers, including magnetoresponsive oil-based ferrofluids. Moreover, we extended the technique to triple-layered encapsulation by overlaying a second interfacial layer atop the first floating interfacial layer.

*Findings:* The encapsulating layer(s) effectively protects the water-soluble inner core (ethylene glycol) inside the water bath. A non-dimensional experimental regime is established for successful encapsulation in terms of the impact kinetic energy, interfacial layer thickness, and the viscosity ratio of the interfacial layer and the outer core liquid. Using selective fluorescent tagging, we confirm the presence of individual shell layers wrapped around the core, which presents a promising pathway to visualize the internal morphology of final encapsulated droplets.

## 1. Introduction

Encapsulation enables the wrapping of the material of interest by another material. The material of interest is termed "core," and the wrapping material is termed "shell." Owing to its unique core-shell morphology and the high flexibility of the material selection, encapsulated cargos could be imparted with diverse properties and functionalities such as controlled release [1], mass transfer [2], and mechanical response [3]. Therefore, they have been extensively utilized in various applications, including agriculture [4], food processing [5], aromatherapy [6], personal care [7], electronics [8] and pharma/nutraceuticals [9]. Encapsulation serves several purposes, such as shielding a vulnerable core from the surrounding conditions (e.g., oxidation and evaporation), protecting the environment from toxic products during its handling, masking desired properties (e.g., taste and odor) of the core material, separating two or multiple reactive substances. Depending on the desired application, the core material in an encapsulation process can exist in different physical states, including solid, gas, or liquid. The shell material can be solid or liquid and should manifest desirable attributes such as stability, strength, flexibility, and impermeability [10,11].

The encapsulation process can produce single or multi-layer encapsulated cargo. In single-layer encapsulation, the core material is wrapped by a single shell layer. Multi-layered encapsulation is achieved by adding one or more wrapping layers between the core and the outermost shell. Note that we define n-layered encapsulation as a process of wrapping a target core by n-shell layers, which is not to be confused with $n^{th}$-order emulsification in microfluidics where the core is also considered an entity in defining the order and, therefore, corresponds to a system where a core drop is wrapped with (n-1) shell-forming layers. According to our definition, a double-layered encapsulated cargo comprises a single core wrapped by two shell layers. Similarly, a triple-layered encapsulated cargo comprises a single core wrapped by three shell layers. Multi-layered cargos provide better protection to the core and enable additional functionalities toward various applications related to bio, food, and pharmaceutical engineering [12–14]. In this regard, the controlled release of active agents like drugs is of significant interest in developing advanced delivery vehicles [15,16]. Multi-layered capsules allow more active agents to be locally and separately embedded in the capsules than single-layered capsules. For example, using the multi-layered encapsulated cargo, various drugs could be embedded in the capsule's core or shells to study the synergistic effects [17]. Depending on the required application, various active or passive methods could be employed for the targeted release of the drugs [18,19]. For example, in the case of a double-layered encapsulation, the outermost shell can be functionalized using magnetic nanoparticles to release the drugs embedded in the core and the middle shell in a controlled fashion using magnet-assisted techniques [20]. Similarly, it is well known that pH is one of the most important target stimuli for triggered release in physiological applications, as pH varies in the human body. A double-layered encapsulated cargo can be fabricated to rupture the outermost shell as it reaches the stomach (pH ~ 2), releasing the embedded drug. Similarly, as the cargo traverses inside the body, the inner shell ruptures as it reaches the intestine (pH ~ 9), releasing drugs from the core [21].

We can categorize encapsulation techniques into physical and chemical methods. Physical methods include coacervation, ionic gelation, solvent evaporation, sol-gel method, spray drying, and electrospraying; chemical methods comprise in situ polymerization, interfacial polymerization, and emulsion polymerization [5,22,23]. However, these methods encounter limitations, including challenges in controlling the thickness of the shell, low yield and stability, and using large numbers of equipment, restricting the utility of these conventional methods. More importantly, these techniques pose significant challenges in realizing multi-layered encapsulated cargos. On the other hand, microfluid-based methods are based on emulsification technology, where the droplets are generated using shear forces [24–30]. Although these techniques offer precise drop size control, they pose several challenges when generating multi-layered encapsulated cargos. In the case of multi-layered encapsulation, the involvement of multiple liquids demands multiple syringe pumps for fluid infusion with accurate flow control [13]. As the number of shell layers increases, the microfluidic device should consist of an array of drop makers with the required wettability contrast at each channel wall. This demands a complex fabrication process [17]. In addition, as the drop generators increase, synchronization of drop formation becomes challenging. If the timing of one of the drop generators in the array fails, the complete encapsulation process is affected [13]. Microfluidic encapsulation using a viscous outer layer presents challenges due to the substantial viscous resistance from the shell liquid and the risk of leakage. Microfluidic techniques find more applications where precision of capsule size is crucial, e.g., drug delivery, than in industrial applications where achieving a high throughput is of primary importance [31–33].

We previously achieved stable, ultrafast, and robust liquid-liquid encapsulation, where a liquid droplet impacts another liquid layer floating on a water bath, giving rise to single-layer encapsulation [34–37]. A liquid core and shell enhance the bioavailability and dosage efficiency of the resulting encapsulated cargo. In a recent study, researchers utilized our patented impact-driven liquid-liquid encapsulation method [37] to create stable double-layered capsules. These capsules allowed subsequent extraction of the encapsulated cargo by UV-curing the outermost shell [38]. It is to be noted in their work [38] that they refer to this type of encapsulation as triple-layered encapsulation by considering the core (glycerol) and the two interfacial layers (silicone oil and 1,6-Hexanediol Diacrylate (HDDA)) as separate layers. In our current framework, this would be referred to as double-layered encapsulation. However, the sole dependence of encapsulation on the impact-driven technique possesses several restrictions to produce multi-layer encapsulated cargos. First, for double-layered encapsulation, it is a prerequisite that the liquid triplet (the host bath and two shell-forming liquids) should always satisfy the criterion for density stratification and interfacial tension, which ensures the floating of two interfacial layers on the host bath. This criterion restricts the choice of compatible liquids for the shell layers for the triple-layered encapsulation. In the study [38], one can argue that if another interfacial layer was dispensed on two existing floating interfacial layers, then such triple-layered encapsulation is possible. However, owing to the challenges and complexities of attaining layering order of the interfacial liquids due to thermophysical incompatibility (i.e., adverse density difference or interfacial tension values) between multiple interfacial layers, a higher possibility of interfacial trapping and consequent restrictive limit on minimum usable core droplet size and impact height might arise. In other words, if the application demands a particular shell layer that cannot float on another interfacial layer or the host bath, this technique possesses restriction. Second, for successful encapsulation of the core drop via complete penetration, the impact-driven approach depends on the kinetic energy of the core drop. To achieve successful encapsulation, the method requires a larger core drop or a higher impact height so that the core drop has sufficient kinetic energy to penetrate through the interfacial layers. The kinetic energy of the impacting core droplet is proportional to the mass of the droplet and the squared impact velocity. In our approach, the droplet achieves the impact velocity solely due to the conversion of its potential energy during gravity-assisted downfall from a vertical height, which means that the squared velocity is proportional to the vertical impact height. Consequently, in the impact-driven approach, below a minimum droplet size and impact height, the drop cannot possess sufficient kinetic energy for successful penetration, which leads to restrictive minimum size and impact height criteria for successful interfacial-penetration-driven encapsulation. In one of our recent works [35], we have shown that when a liquid drop of smaller size or possessing insufficient kinetic energy impacts an interfacial layer of poly(dimethylsiloxane) (PDMS) floating on the host water bath, it gets trapped at the interfacial layer. We showed that the interfacially trapped state is also an encapsulated state, given that the interfacial liquid has a positive spreading coefficient on the oncoming core drop, and we also show a strategy to harness the potential of the interfacially trapped state via shell-hardening and subsequent extraction of the encapsulated cargo. However, the interfacially trapped state might be of limited utility in some applications. For example, in applications such as drug targeting and encapsulated flavours in packaged beverages, creating standalone encapsulated cargo that remains separate from the interfacial layer becomes essential. Additionally, wrapping water-soluble cores within a water host bath using the original impact-driven liquid-liquid encapsulation technique poses challenges to the dissolution of the core cargo during the process and consequently leads to a lower success rate of encapsulation.

The present work develops a novel approach to liquid-liquid encapsulation by combining a Y-junction and our impact-driven liquid-liquid encapsulation technique in multi-layer encapsulations. In the following sections, we showcase how this novel technique effectively addresses the challenges mentioned above while maintaining the simplicity and robustness of our previous approach. In this technique, the Y-junction generates compound droplets consisting of an inner and outer core. These compound droplets generated from the Y-junction impact the single or double interfacial layers floating on the host bath, generating double-layered or triple-layered encapsulated cargos. This technique is further exploited to encapsulate water-soluble analytes inside liquid shells, which is challenging by the impact-driven liquid-liquid encapsulation technique. Further, we have utilized this hybrid liquid-liquid encapsulation technique to generate water-in-oil-in-oil (W/O/O) triple-layered encapsulated cargo inside a host water bath.

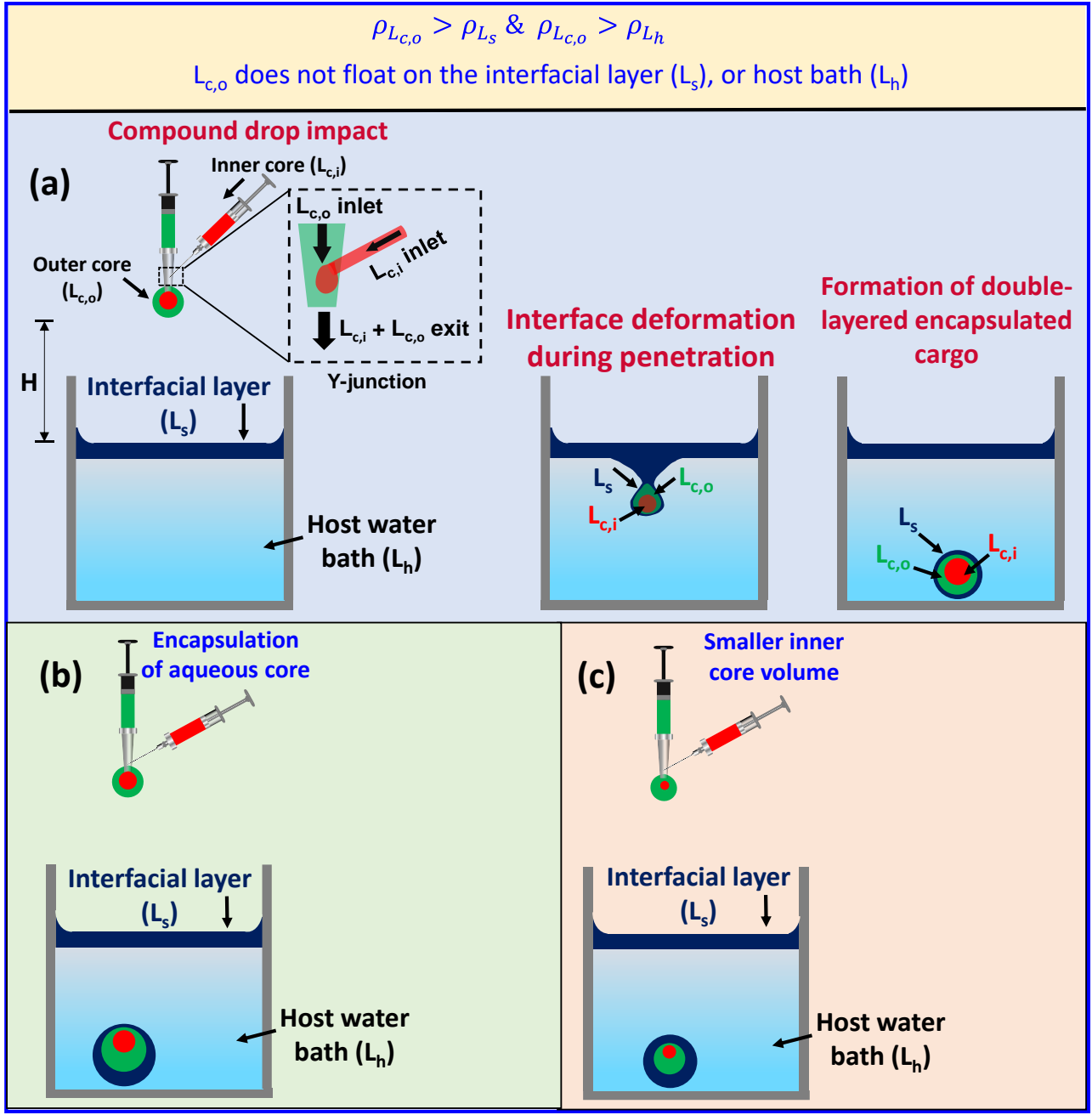

**Figure 1:** (a) The schematic of the hybrid liquid-liquid encapsulation framework for double-layered encapsulation, which involves impact-driven wrapping of a compound droplet generated by Y-junction geometry containing the outer core liquid ($L_{c,o}$) and the inner core liquid ($L_{c,i}$) represented using green and red color, respectively. The interfacial layer ($L_s$) floating on the host water bath (blue) is represented using the dark blue color. (b) Stable wrapping of an aqueous core liquid (red) by two oil layers ($L_{c,o}$ and $L_s$) inside a host water bath. (c) Impact-driven formation of double-layer encapsulated droplet with a small inner core drop (red) inside the two oil layers $L_{c,o}$ and $L_s$. The outer

core liquid ($L_{c,o}$) is heavier than all the liquids involved in all the cases. The impact height ($H$) shown in (a) represents the gap between the dispensing nozzle and the interfacial layer. The density of the outer core liquid, the inner core liquid, and the host bath are as follows: $\rho_{L_{c,o}}$, $\rho_{L_{c,i}}$ and $\rho_{L_h}$, respectively.

## 2. Experimental

**Materials**

The framework involves multi-layered encapsulation; thus, the liquids are named in a particular manner, as shown in Figure 1a. In the case of double-layered encapsulation (Figure 1a), the liquids in the Y-junction involve the inner core and the outer core liquids, which are termed $L_{c,i}$ and $L_{c,o}$, respectively. The subscripts "c,i" and "c,o" indicate the inner and outer core, respectively. The host liquid, $L_h$ is chosen to be deionized (DI) water purified by Milli-Q, MilliPoreSigma, Ontario, Canada) with density $\rho_h = 1000 \text{ kg/m}^3$, dynamic viscosity $\mu_h = 1 \text{ mPa s}$, liquid-air surface tension $\gamma_h = 72 \text{ mN}/m$. The liquid that floats on the host bath is termed the interfacial layer $L_s$, forming the outermost shell, as indicated in Figure 1a. This nomenclature is followed for all the cases shown in Figures 1a to 1c. Throughout the study, the outer core liquid ($L_{c,o}$ in the Y-junction) is kept fixed, which is a class of laser liquid – a mixture of silicanes and polyphenol ethers with a water solubility of $< 0.1 \%$ (Product Code: 57B63, Cargille Laboratories Inc., Cedar Grove, NJ, USA). The relevant material properties are as follows: density $\rho_{c,o} = 1900 \text{ kg/m}^3$, dynamic viscosity $\mu_{c,o} = 1024 \text{ mPa s}$, liquid-air surface tension $\gamma_{c,o} = 50 \text{ mN}/m$, and liquid-water interfacial tension $\gamma_{c,o-h} = 39.4 \text{ mN}/m$. The experiments were performed in a distortion-free glass cuvette (Product Code: SC-02, Krüss GmbH, Hamburg, Germany) of inner dimension $36 \text{ mm} \times 36 \text{ mm} \times 30 \text{ mm}$ with a 2.5 mm wall thickness.

In the case of triple-layered encapsulation demonstrated in section 3.4 (Figure 4e), the liquids in the Y-junction (outer and inner core) and the host water bath remain the same as in the case of double-layered encapsulation. We have used two interfacial layers of mineral oil ($L_{s,1}$) and silicone oil ($L_{s,2}$) floating on the host water bath. The silicone oil interfacial liquid ($L_{s,2}$) was dispensed directly on top of the host water bath ($L_h$), while the mineral oil interfacial layer was created by dispensing mineral oil on top of the already floating silicone oil interfacial layer. The combination of silicone and mineral oil has been widely used to form a co-flow in microchannels for studying the droplet and particle migration across the fluid-fluid interface [33,39]. In these previous studies [33,39] and the present study, this combination is found to be stable, and the oils are not partially miscible. The relevant material properties for mineral oil are as follows: density $\rho_{L_{s,1}} = 877 \text{ kg/m}^3$, dynamic viscosity $\mu_{L_{s,1}} = 28.7 \text{ mPa s}$, liquid-air surface tension $\gamma_{L_{s,1}} = 28.09 \text{ mN}/m$, liquid-water interfacial tension $\gamma_{L_{s,1}-L_h} = 25\,\mathrm{mN}/m$. The relevant material properties for silicone oil are as follows: density $\rho_{L_{s,2}} = 1090\,\mathrm{kg/m^3}$, dynamic viscosity $\mu_{L_{s,2}} = 1000\,\mathrm{mPa\,s}$, liquid-air surface tension $\gamma_{L_{s,2}} = 27.77\,\mathrm{mN}/m$, liquid-water interfacial tension $\gamma_{L_{s,2}-L_h} = 35.7\,\mathrm{mN}/m$. The interfacial tension [39] between the silicone oil and the mineral oil is $\gamma_{L_{s,1}-L_{s,2}} = 1.11 \pm 0.1\,\mathrm{mN}/m$.

**Methods**

Prior to each experiment, the glass cuvette undergoes a thorough cleaning process. It is soaked and subjected to ultrasonication (using the Branson 5800 from Emerson Electric Co., USA) in hexane for 30 minutes. Afterward, the cuvette is rinsed with deionized water and acetone, followed by drying with compressed nitrogen. Subsequently, the cleaned cuvette is treated with air plasma (using the PE-25 from PLASMA ETCH, USA) for 10 minutes. Finally, the cuvette is positioned on a vertically movable stage provided by Kruss GmbH in Hamburg, Germany. Initially, the cuvette was partially filled with 20 ml of the host liquid (DI water). Next, a predetermined volume of the interfacial liquid was dispensed onto the water bath from close proximity using a pipette (DiaPETTE, Canada). The interfacial liquid was allowed to spread uniformly for ~ 2 min, forming a thin interfacial layer (which acts as the shell layer) with a specific thickness. Finally, the dispenser was moved to the predetermined height ($H$) using the linear translating stage. The impact height in the case of double-layered encapsulation varies between $H = 5$ cm to $H = 46$ cm. In Figure 1a, we observe the impact of a compound droplet on the interfacial layer. A Y-junction flow arrangement was employed to create the compound droplet. Laser oil was pushed through a vertically oriented microtip (with an inner diameter of 2 mm) attached to a syringe. Simultaneously, the core drop was introduced from the side using a flat-tipped stainless-steel needle (gauge 25, part no. 7018339, Nordson EFD, USA) with an internal diameter of 0.25 mm, mounted on a 1 ml NORM-JECT syringe. The compound droplet was then dispensed by pumping laser oil using the programmable syringe pump (Chemyx Fusion 4000) at a controlled rate. After detaching from the microtip, the compound droplet accelerates downward under the influence of gravity and encounters the interfacial layer. Owing to sufficient kinetic energy, the compound drop successfully surmounts both the interfacial and viscous barriers, ultimately penetrating through the interfacial layer. As a result, the interfacial layer wraps around the compound drop. To study the complete dynamics of the encapsulation process, we employed a high-speed camera (Photron FASTCAM Mini AX200) operating at 6400 frames per second (fps). The camera was equipped with a zoom lens (Navitar 7000 Zoom), an effective focal length ranging from 18 mm to 108 mm, and an internal memory of 32GB. The high-speed camera allowed a maximum recording window of approximately 3.4 seconds at full resolution (1024 px×1024 px), sufficient to capture the entire time-resolved encapsulation process (typically completed within 500 ms). The 'end' trigger mode was used for image acquisition, triggered by visually confirming the completion of encapsulation. Subsequent analysis involved transferring the captured data to a personal computer connected to high-speed Gigabit Ethernet. Additionally, color images were captured using a DSLR camera (Nikon D5200) for enhanced visualization

To achieve triple-layered encapsulation, first, a predetermined volume of the silicone oil ($L_{s,2}$) is dispensed on top of the water bath from proximity using a pipette and allowed to spread uniformly for ~ 2 min, resulting in the formation of a thin interfacial layer of respective thickness. Then, the mineral oil ($L_{s,1}$) is dispensed on the floating layer of silicone oil and allowed to spread uniformly on the silicone oil interfacial layer. This process results in double interfacial layers floating on the host water bath. As discussed earlier in this section, the compound droplet generated using the Y-junction geometry impinges on the interfacial layers floating on the host water bath. The compound drop overcomes the interfacial and viscous barriers and penetrates through the interfacial layers due to sufficient kinetic energy. This process results in the wrapping of the compound drop by the two wrapping layers. The detailed experimental demonstration of triple-layered encapsulation is depicted in section 3.4.

## 3. Results and Discussion

This hybrid method of liquid-liquid encapsulation consists of two processes that involve generating a compound droplet using Y-junction and the subsequent impact of the compound droplet on the interfacial layer floating on the host water bath. The Y-junction consists of two arms, one containing the inner core liquid ($L_{c,i}$) and another containing the outer core ($L_{c,o}$), as shown in Figure 1a. Two possibilities arise when the compound droplet is dispensed on the interfacial layer ($L_s$) floating on the host liquid bath from a vertical separation ($H$) (Figure 1a). If the compound drop possesses sufficient kinetic energy, it can overcome the viscous and interfacial resistance of the interfacial layer, leading to penetration through that layer. Conversely, if the impact kinetic energy is dissipated due to viscous resistance before the compound droplet can detach from the interfacial layer, the compound droplet gets trapped at the interfacial layer. To quantify the impact kinetic energy of the compound droplet, we have used the impact Weber number ($We_i$). The impact Weber number is defined as $We_i = [\beta\rho_{c,i} + (1-\beta)\rho_{c,o}]\frac{D_c V^2}{\sigma_{c,o}}$, where $\beta$ is the volume fraction of the inner core in the compound droplet, $\rho_{c,i}$ and $\rho_{c,o}$ are the densities of the inner core ($L_{c,i}$) and outer core ($L_{c,o}$), respectively. $D_c$, $\sigma_{c,o}$ are the diameter of the compound drop and the surface tension of the outer core liquid. $V$ is the impact velocity of the compound droplet, estimated as $V = \sqrt{2gH}$, where $g$ is the acceleration due to gravity, $H$ is the impact height.

### 3.1. Underlying mechanism involved in the double-layered encapsulation

Figure 2 and Supporting Movie 1 depict the detailed visual representation of the interface dynamics during the encapsulation process of the compound core droplet. This intricate process involves a dynamic interplay between interfacial tension forces, viscous forces, and momentum at the five-fluid interface (involving air, the interfacial layer, host liquid, outer core, and inner core). The process leads to the formation of a double-layered encapsulated droplet at the bottom of the cuvette, which can be seen in the final two snippets (the high-speed image corresponding to $t = 56.40\ ms$ and the final color image of the settled droplet) of Figure 2. Owing to the vertical separation $H = 7\ cm$ between the interfacial layer ($L_s$) and the dispensing needle, the core drop gains a kinetic energy before it encounters the interfacial layer ($t = 0.00\ ms$). The compound drop possessing sufficient kinetic energy, encounters the interfacial layer ($t = 11.71\ ms$). Upon contact with the interfacial layer ($L_s$), the compound drop attempts to penetrate through the interfacial layer ($L_s$), as indicated by "interfacial penetration by the compound drop" in Figure 2. The compound drop drags the $L_s$ layer downward through the host liquid ($L_h$) due to its momentum, which deforms the interfacial layer-water interface and increases its surface area. Despite the downward motion of the drop, interfacial forces acting on the deformed interface between $L_s$-$L_h$ strive to restore it to its original position, minimizing interfacial energy. Additionally, the viscous resistance from the interfacial layer opposes the drop's motion by dissipating its momentum. This competition leads to the formation of the neck, and interfacial penetration continues till $t = 16.56\ ms$. Owing to the sufficient momentum, the compound drop overcomes the barrier imposed by both the interfacial forces and the viscous resistance, leading to the initiation of the necking process ($t = 24.21\ ms$), which continues till $t = 31.40\ ms$. During the necking of the interfacial layer (canola oil), the thinning of the liquid threads of canola oil can be seen as indicated in Figure 2. During the process, the droplet travels toward the bottom of the cuvette. Finally, the neck thins beyond a critical thickness ($t = 38.11\ ms$), resulting in the separation of the wrapped compound droplet from the interfacial layer ($t = 38.11\ ms$). Subsequently, the interface returns to its before-impact unperturbed position. As the detached layer wraps around the core compound drop, it forms a thin enclosure known as the encapsulating shell. Finally, the wrapped compound droplet settles at the bottom of the cuvette ($t = 56.40\ ms$). During the complete process of double-layered encapsulation, the inner core drop consisting of water-soluble ethylene glycol (EG) remains stable, as seen in Figure 2. The rightmost color image in Figure 2 depicts the final settled wrapped cargo, which consists of the inner EG drop wrapped by two wrapping layers: the laser oil and the canola oil.

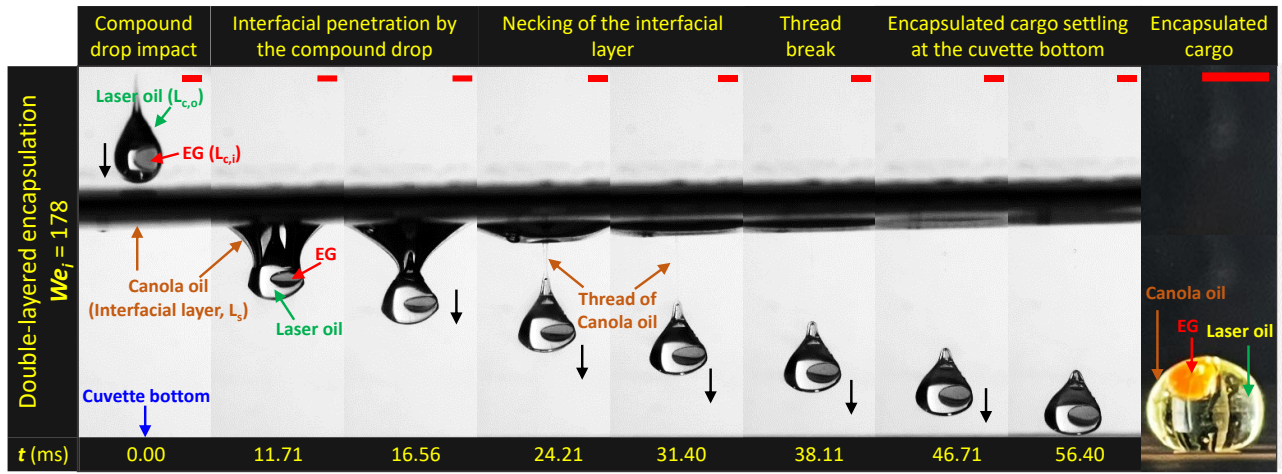

**Figure 2:** Time-resolved high-speed image sequence of double-layered encapsulation showing the interface evolution during the impact-driven wrapping of EG-laser oil compound core droplet (generated via Y-junction as shown in Fig. 1a) by canola oil interfacial layer. The various steps of the encapsulation process are represented, and the liquids involved are identified with different colors. The cuvette bottom is indicated using the blue colored arrow. The motion of the droplet is indicated using a black arrow. The impact Weber number associated with the encapsulation process is $We_i = 178$ (achieved using $H = 7$ cm). The rightmost color image depicts the final settled wrapped cargo. The time is in milliseconds (ms), and the scale bar in each image represents 2 mm.

### 3.2. Thermodynamic criteria for the successful encapsulation

In our previous studies [34–37] on encapsulating a single core droplet by an interfacial layer floating on a host water bath, we established the thermodynamic criterion for forming a stable encapsulated droplet. In those studies, laser oil was used as a core droplet, and canola oil was used as the interfacial layer floating on the host water bath. The same criterion can be extended to the present system as follows by suitably substituting the interfacial tension values of the participating fluids:

$$\gamma_{L_{c,o}-h} - \gamma_{L_{c,o}-L_s} - \gamma_{L_s-L_h} > 0 \tag{1}$$

The equation above relies on minimizing interfacial energy before and after impact, depending solely on surface and interfacial tensions. Note that the surface/interfacial energy of the inner core ($L_{c,i}$) is not important in determining the thermodynamic favorability of impact-driven encapsulation of the compound droplet. The inner core is already enclosed by the outer core ($L_{c,o}$, here laser oil) in the compound droplet. Therefore, it does not participate in interfacial interactions with the shell liquid ($L_s$) or the host bath ($L_h$). Instead, the interfacial energy of the outer core is relevant for considering the overall energetic favorability of the final double-layered encapsulated droplet. The criteria presented in (1) are formulated accordingly. Suppose we substitute the interfacial tension values of liquid used in our experiments mentioned in (1). In that case, we get that for our combination involving laser oil (core), canola oil (shell), and water (host liquid), yields $(39.4 - 2.22 - 18.01)$ mN/$m = 19.17$ mN/$m > 0$.

Based on the experimental images and the criteria mentioned, we can confirm that the compound droplet remains stably encapsulated within the canola oil. Similarly, when considering an oil-based ferrofluid as the interfacial layer (shell) liquid, we find that the interfacial tension between the shell layer and laser oil is $4\,\mathrm{mN}/m$, while the interfacial tension between the shell layer and the host water bath is $23.6\,\mathrm{mN}/m$ [36]. Using these values, we establish that $(39.4 - 4 - 23.6)\,\mathrm{mN}/m = 11.8\,\mathrm{mN}/m > 0$. This result indicates that the thermodynamic criteria are satisfied for this specific liquid-liquid combination. The ethylene glycol (EG) droplet (marked with the red arrow in Figure 2) remains stable during interfacial evolution, as observed in the high-speed images. Furthermore, the EG droplet maintains stability inside the host water bath after encapsulation.

### 3.3. Regime map for double-layered encapsulation

While thermodynamic favorability provides insight into the equilibrium configuration of the encapsulated state, it doesn't fully account for the irreversible nonequilibrium processes during the interaction between the core droplet and the interfacial layer. These dynamic processes play a crucial role in determining the overall behaviour of the compound droplet. Figure 3a shows the non-dimensional regime map depicting the successful encapsulation of the compound droplet with the variation in the viscosity of the interfacial layer by using three different interfacial liquids- canola oil, silicone oil and PDMS. Figures 3b and 3c depict the non-dimensional regime map for the variation of interfacial layer thickness for the canola oil and the silicone oil, respectively. In our study, we focus on the essential parameters that determine the formation of double-layered encapsulated cargo. Specifically, we analyze the impact energy of the compound droplets and the thickness of the interfacial layer. The interplay between the momentum of the impacting compound droplet, the viscous resistance from the interfacial layer, and the restorative interfacial forces at the deformed $\mathrm{L_s}$-$\mathrm{L_h}$ interface ultimately governs the outcome of the impact. The result indicates whether the compound droplet will be able to penetrate through the interfacial layer or will be trapped at the interfacial layer. In both cases, the compound droplet will be wrapped by the interfacial layer, as shown in one of our recent studies [35]. Experiments with various liquids as interfacial layers ($\mathrm{L_s}$) of different viscosity and thickness are performed by varying the impact energy of the compound droplet. In Figure 3a, a non-dimensional experimental regime for encapsulation is identified in terms of the impact Weber number $We_i$, and the viscosity ratio $\alpha$. The viscosity ratio is defined as the ratio of the viscosities of the interfacial layer ($\mathrm{L_s}$) and the outer core ($\mathrm{L_{c,o}}$) as $\alpha = \frac{\mu_{L_s}}{\mu_{L_{c,o}}}$. In Figure 3a, the volume of each of the interfacial layer is kept fixed at $V_{L_s} = 120\,\mu\mathrm{L}$. The impact Weber number $We_i$ is varied over a broad range by changing the impact height $H$ between the interfacial layer and the dispensing needle and the volume fraction of the inner core in the compound droplet $\beta$. In our experiments, $We_i$ for canola oil is varied between 127 (corresponding $H = 5\ cm$) to 280 (corresponding $H = 11\ cm$), for silicone oil is varied between 152 (corresponding $H = 6\ cm$) to 763 (corresponding $H = 30\ cm$) and for PDMS, the variation in $We_i$ is 152 (corresponding $H = 6\ cm$) to 1171 (corresponding $H = 46\ cm$). In Figure 3a, for canola oil ($\alpha = 0.062$), the transition from interfacial trapping to encapsulation via complete penetration occurs at $We_i = 165$. On the contrary, for PDMS ($\alpha = 3.41$), the transition occurs at a higher $We_i = 611$. This fact can be attributed to the higher viscosity of the PDMS, which is 55 times that of the canola oil's viscosity. The high viscosity of the PDMS interfacial layer results in a substantial increase in viscous resistance. Consequently, the transition shifts from interfacial trapping to penetration at a higher Weber number $We_i$ compared to canola oil, even when the same volume $V_{L_s}$ is considered. In Figures 3b and 3c, a non-dimensional experimental regime for encapsulation is identified for canola oil and silicone oil in terms of the impact Weber number, $We_i$, and the non-dimensional interfacial film thickness ($\delta^*$), obtained by non-dimensionalizing the interfacial layer thickness with respect to the compound droplet diameter. The primary variation of $We_i$ is achieved by varying the impact height $H$ mentioned above. However, to depict the variation of $We_i$ with the variation of $\beta$, we have performed some experiments where the diameter of the inner core (EG droplet) is varied from 1.51 mm to 2.94 mm. The compound droplet diameter is kept fixed at 3.46 mm. The resulting variation in $\beta$ is from 0.076 to 0.38, respectively. In Figures 3b and 3c, the volume of canola oil is varied from $V_{L_s} = 120\ \mu L$ to 420 $\mu L$, and the volume of silicone oil is varied from $V_{L_s} = 30\ \mu L$ to 180 $\mu L$. When dispensed on the water-air interface, canola oil forms a floating bi-convex oil lens with one side in contact with water while the other is exposed to air. The thickness of the canola oil is estimated using the methodology used in our previous study [34]. However, silicone oil, upon dispensing on the host water bath, results in the formation of a uniform dimensional thickness estimated as $\delta = \frac{V_{L_s}}{A_c}$, where $A_c$ is the inner cross-sectional area of the cuvette [35]. In our experiments, $We_i$ for canola oil is varied between 127 to 280, that corresponds to the height variation from $H = 5\ cm$ to $H = 11\ cm$. In case of silicone oil, $We_i$ is varied between 152 to 763, that corresponds to the height variation from $H = 6\ cm$ to $H = 30\ cm$. Based on our experiments, we observed that the dynamics of compound droplet encapsulation remain similar for each interfacial layer. In Figure 3b, when canola oil is used as the interfacial layer, and its thickness is reduced, a compound droplet with lower impact kinetic energy can still penetrate the interfacial layer. This behaviour occurs because the thinner interfacial layer offers lower viscous resistance. Therefore, successful encapsulation is achieved upon impact, and the encapsulated cargo forms inside the host bath. On the other hand, when the thickness of the canola oil is increased, a higher impact kinetic energy (higher $We_i$) of the compound droplet is desired for the compound droplet to penetrate through the interfacial layer, resulting in the encapsulated cargo forming inside the host bath. Consequently, a gradual transition from the trapped state of encapsulation to the encapsulation through complete penetration could be observed in the regime map presented in Figure 3b, with an increase in $\delta^*$. The transition is identified using the $We_i$ which indicates whether the compound drop will be trapped or will be able to penetrate through the interfacial layer. Similarly, when the silicon oil (Figure 3c) is used as an interfacial layer, it can be observed that the $We_i$ at which transition from the trapped state of encapsulation to the encapsulation through complete penetration of the interfacial layer is higher, even for a lower $\delta^*$ as compared to the canola oil. The difference in viscosity between silicone oil ($\approx 16$ times higher than canola oil) and canola oil plays a crucial role. Due to its high viscosity, silicone oil offers significantly greater viscous resistance. As a result, interfacial trapping occurs even at a much lower interfacial layer thickness.

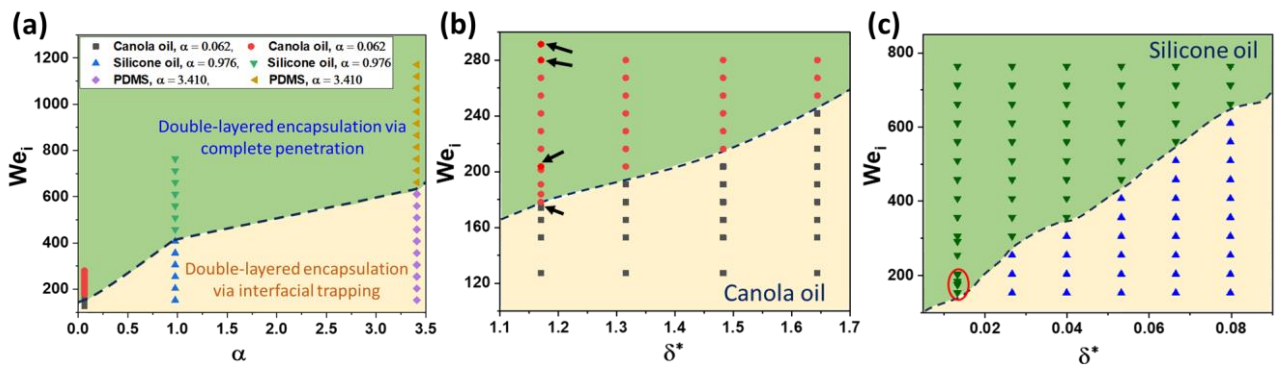

**Figure 3:** (a) Regime map depicting the region for the successful double-layered encapsulation via interfacial trapping and encapsulation via complete penetration for canola oil, silicone oil, and PDMS as the interfacial layers, respectively. (b-c) Depicts the relationship between the non-dimensional thickness of the interfacial layer and the impact Weber number for achieving successful double-layered encapsulation through either interfacial trapping or complete penetration for canola oil and silicone oil interfacial layers, respectively. The black arrows in (b) and the encircled region (red) in (c) represent the experimental data corresponding to the $We_i$ when the volume fraction $\beta$ of the inner core in the compound droplet is varied, keeping the height of impact fixed $H = 6$ cm for canola oil and $H = 7.5$ cm for silicone oil. Here, $\alpha$ is the viscosity ratio defined as $\alpha = \frac{\mu_{L_s}}{\mu_{L_{c,o}}}$, where $\mu_{L_s}$ is the viscosity of the interfacial layer liquid ($L_s$) and $\mu_{L_{c,o}}$ is the viscosity of the outer core liquid ($L_{c,o}$). Here $\delta^*$ represents the non-dimensional interfacial film thickness obtained by non-dimensionalizing the interfacial layer thickness with respect to the compound core droplet diameter. In the case of (a), different symbols are used for a particular interfacial layer liquid to demarcate the outcome of encapsulation (via complete penetration and via interfacial trapping). The upper green regime in the case of (a), (b), and (c) represents encapsulation via complete penetration, and the yellow regime represents encapsulation via interfacial trapping.

## 3.4. Experimental images of various test cases performed using the hybrid liquid-liquid encapsulation technique

The practical significance of efficient encapsulation lies in protecting target analytes by shielding them from harsh environments and preventing unintended release. Figure 4e illustrates how the hybrid technique framework can successfully generate double-layered encapsulated cargos (as shown in Figure 4a) within a water bath. In this regard, we have used a compound droplet consisting of ethylene glycol (EG) as the inner core liquid ($L_{c,i}$) and laser oil as the outer core ($L_{c,o}$), as shown in Figures 4b, 4c and 4d. We have used three different interfacial layers ($L_s$) - dyed canola oil of volume $V_{L_s} = 120$ µL (Fig. 4b), silicone oil of volume $V_{L_s} = 90$ µL (Fig. 4c), and oil-based ferrofluid of volume $V_{L_s} = 200$ µL (Fig. 4d) as the interfacial layer ($L_s$) for double-layered encapsulation. For triple-layered encapsulation (Fig. 4e), we have used two different interfacial layers ($L_s$) - mineral oil ($L_{s,1}$, Fig. 4f) and silicone oil ($L_{s,2}$, Fig. 4g). The selection of EG as one of the core liquids shows that despite being water-soluble, our novel hybrid encapsulation technique can provide efficient protection of the encapsulated cargo inside the host water bath.

A demonstrative color video is presented in Supporting Movie 2 of the double-layered encapsulation process with canola oil (dyed red) as the interfacial layer. We also depicted a test case using an oil-based ferrofluid as the interfacial layer, which offers several functional advantages, including magnet-assisted manipulation [36,40–44]. In each case, the choice of $We_i$ is based on the regime map depicted in Figure 3. This ensures that the impinging compound droplets possess sufficient kinetic energy to penetrate through the interfacial layer. Once successful penetration occurs, all the droplets are enveloped by their respective interfacial layers, as schematically shown in Figure 4a. Furthermore, we confirm that despite the thinness of the wrapping layer, the encapsulated cargo remains stable in water. There is no evidence of the miscible inner core perishing or dissolution ithin the water environment. Further, we show the encapsulation of the compound droplets by two wrapping layers, i.e., mineral and silicone oil floating on the host water bath. This scenario bolsters our hybrid technique by demonstrating a robust and simple technique to achieve triple-layered encapsulation. Figures 4e-g depict the triple-layered encapsulation of compound droplets consisting of EG as the inner core ($L_{c,i}$) and laser oil ($L_{c,o}$) as the outer core. The compound droplets impact the two interfacial layers (mineral oil and silicone oil) floating on the host water bath. The silicone oil interfacial liquid ($L_{s,2}$) was dispensed directly on top of the host water bath ($L_h$), while the mineral oil interfacial layer was created by dispensing mineral oil ($V_{L_{s,1}} = 120$ µL) on top of the already floating silicone oil interfacial layer ($V_{L_{s,2}} = 90$ µL), as can be seen in Figure 4(e). The $We_i$ is chosen in such a manner that it ensures the complete penetration of the compound droplet through the two interfacial layers, resulting in the formation of the triple-layered encapsulated cargo inside the host bath. As shown in the schematic of Figure 4e, in the final encapsulated droplets (Figures 4f, g), the inner core is wrapped by three oil layers. As we move radially outward, starting from the EG inner core ($L_{c,i}$), we first encounter the laser oil outer core ($L_{c,o}$), followed by an intermediate wrapping layer of mineral oil ($L_{s,1}$) and a final outer wrapping layer of silicone oil ($L_{s,2}$). Note that optical visualization and the distinction between the wrapping layers in the encapsulated droplets become challenging for multi-layered encapsulation. In the case of triple-layered encapsulation (Figures 4f and 4g), we have used a blue oil-based dye to stain the two interfacial layers individually, one at a time, for distinctive visualization. In Figure 4f, the intermediate shell-forming liquid, mineral oil ($L_{s,1}$), is dyed blue, while in Figure 4g, the blue dye is added to the outer shell-forming liquid, silicone oil ($L_{s,2}$). The blue dye trace is present in both cases (Figures 4f, g), and it is more prominent when the dye is applied to the outer shell liquid (Figure 4g). A visual comparison of the color intensity of the blue dye trace indirectly confirms successful wrapping by two interfacial layers with the desired layer sequence. The colored visualization of the triple-layered encapsulation process involving blue-dyed mineral oil and silicone oil as interfacial layers can be seen in Supporting Movie 3, as shown in Figure 4e. Also, please note that the accumulation of the shell-forming liquid near the apex is due to the buoyancy-driven upward migration of the excess shell fluid, which is lighter in density than the host bath. Although the excess interfacial layer forming liquid tends to move upward, an all-around wrapping film remains intact around the core, as dictated by equilibrium thermodynamics [45,46].

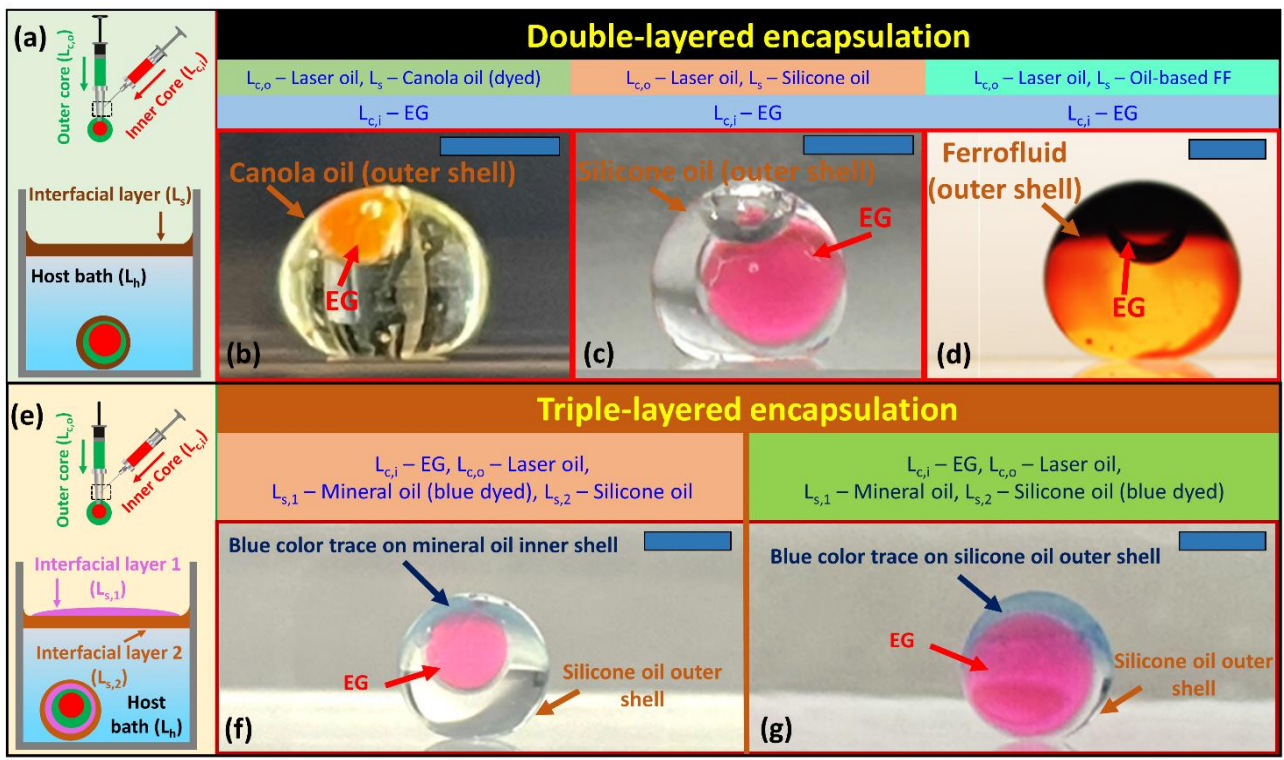

**Figure 4:** Demonstration of various use cases showing multi-layered encapsulation using the hybrid liquid-liquid encapsulation technique. (a) Schematic representation of the hybrid double-layered liquid−liquid encapsulation framework leading to the formation of wrapped droplets at the bottom of the cuvette – (b) Canola oil (dyed) as the interfacial layer ($We_i = 178$, $H = 7$ cm), (c) Silicone oil as the interfacial layer ($We_i = 356$, $H = 14$ cm), (d) Oil-based ferrofluid as the interfacial layer ($We_i = 127$, $H = 5$ cm). (e) Schematic representation of the hybrid triple-layered liquid−liquid encapsulation framework leading to the formation of wrapped droplets at the bottom of the cuvette – (f) Dyed mineral oil ($L_{s,1}$) and silicone oil ($L_{s,2}$) are used as the interfacial layer, (g) Mineral oil ($L_{s,1}$) and dyed silicone oil ($L_{s,2}$) are used as the interfacial layer. The impact Weber number and the impact height corresponding to the triple-layered encapsulation are ($We_i = 356$, $H = 14$ cm). In all the experiments, laser oil is used as the outer core ($L_{c,o}$), and dyed EG is used as the inner core droplet

(L$_{c,i}$). The scale bar is 2 mm throughout the figure. Note that the inner core size is varied between experiments to emphasize that the technique remains applicable over a broad range of core drop sizes. The relevant volumes of the inner EG core in Figures (b), (c), (d), (f), and (g) are 1.80 μL, 9.41 μL, 3.31 μL, 3.15 μL and 13.30 μL, respectively.

### 3.5. Visualization of wrapping layer using Fluorescence microscopy

As discussed in the previous section, optical visualization of the wrapping layer in the encapsulated cargo is extremely challenging due to the diffraction limit of light. As is the case here, the challenges become even more pronounced when multiple wrapping layers are introduced. In Fig. 4 (f-g), we attempted to visualize the wrapping layer by selectively adding dye to the interfacial layers one at a time. However, this approach to visualization remains subjective. Here, we attempt to confirm the existence of the wrapping layer by fluorescence imaging of the encapsulated droplet with the suitable addition of fluorescent dye to the involved fluidic entities, which allows us to investigate the internal morphology of the encapsulated droplet qualitatively.

First, as a control, we demonstrate the visualization of the wrapping layer in a single-layer encapsulated cargo. We follow the method outlined in our previous works to generate the single-layered encapsulated cargo [34,35]. Instead of using a compound droplet generated from a Y-junction as the core drop, we use a single component laser oil droplet as the core and imping it onto a floating interfacial layer of canola oil. We added a red fluorescent dye (made of Red Fluorescent Polymer microspheres, diameter 1-5 μm, Cospheric Inc.) to canola oil. Once the encapsulated droplet is formed at the bottom of the cuvette, we capture the 3D morphology of the entire droplet by Z-stacking (optical slicing of the droplet along the optical axis of the microscope via automated adjustment of the relative focal plane) in confocal mode using a laser scanning microscope (LSM-800, Carl Zeiss). In this case, we use a 561 nm laser source to raster scan the droplet slice-by-slice. The used red fluorescent tracers have an emission peak at ∼ 607 nm. Therefore, we use a detection range of 597-700 nm to capture the emitted fluorescent signal after they are routed via the confocal pinhole. The relative focus position (i.e., slicing location) with respect to the base plane of the droplet is denoted by $d$, as shown in the schematic representation of the encapsulated droplets in Figure 5a. The coordinate system for denoting slicing directions and section planes is also marked in Figure 5a. In Figure 5b (i), we present several illustrative Z-slices of the single-layered encapsulated droplet that capture the morphological evolution of the droplet as the relative focus positions are shifted upward along the Z-axis. The corresponding $d$ values are indicated on the top left corner of the slices in $\mu m$. These slices essentially represent the X-Y projection of the drop shape captured at a vertical separation $d$ from the base plane of the droplet. Several observations can be made from the series of Z-stack slices presented in Figure 5b (i). First, the inner ring visible in the slice corresponding to $d = 300$ μ$m$ marks the apparent contact surface of the droplet with the bottom substrate. The visible fluorescent signature along the apparent contact surface indicates the presence of the shell-forming liquid along the droplet-surface contact and, therefore, stands in line with our previous observation that reports distinctive similarity in the wetting signature of the encapsulated droplet (inside host bath) with that of the shell-forming liquid [34]. Further, the fluorescent outer rim of the droplet also provides confirmative evidence of the wrapping layer around the outer periphery of the droplet. As expected from the geometry of the encapsulated droplet (Figure 5a), the diameter of the outer rim increases as $d$ increases until we reach the widest section of the drop (corresponding projected diameter, $D_{proj}^{max}$, see Figure 5a) on XZ plane, and then it reduces. For the case demonstrated in Figure 5b (i), $d = 1300 \ \mu m$ corresponds to the occurrence of $D_{proj}^{max}$. Next, as the relative focus plane moves upward beyond $D_{proj}^{max}$, a noticeable increase in the fluorescent intensity is observed in the captured slices. This can be attributed to the buoyancy-driven upward migration of the excess interfacial liquid in the shell layer and eventual accumulation near the apex of the wrapped droplet, a phenomenon that we reported in our previous works [34,36]. Note that the acquired slices can be used to construct the 3D shape of the droplet and also orthographic projections along different sectioning planes. In Figure 5c (i-iii), we present one such orthographic projection. For XZ (Figure 5c (i)) and YZ (Figure 5c (iii)) projections, vertical cut sections along the central plane of the droplet were taken, while for XY (Figure 5c (ii)) projection, a horizontal section along $D_{proj}^{max}$ was taken. The outward expansion of the outer rim can also be captured in the XZ projection presented in Figure 5c (i). The side view of the single-layered encapsulated cargo is presented in Figure 5c (iv). The upward accumulation of the red-dyed interfacial layer can also be confirmed from the side view.

It is to be noted that during the impact-driven interfacial penetration, depending on the kinetic energy of the core droplet and the timescale of necking of the interfacial layer, it is also possible to entrap the surrounding air phase in the encapsulated droplet. For example, when the core droplet has high kinetic energy (i.e., high impact height $H$) or the viscous resistance offered by the interfacial layer is low (i.e., lower interfacial layer thickness or low viscosity of the interfacial fluid), the impinging core drop can drag the interfacial layer downward through the host bath much faster than the neck could thin down, creating space for the surrounding air to be trapped within when the neck eventually closes. The encapsulated cargo separates from the interfacial layer. To showcase the versatility of our technique, we have internationally included the trapped air phase for both the single-layered and the double-layered encapsulated droplets presented in Fig. 5. Entrapment of the gas phase outside the interfacial (shell) layers could of interest to several engineering applications, where post-encapsulation, one can achieve a micro-reactor with the appropriate choice of the surrounding gas phase (albeit inert and inviscid air phase used in the experiments). However, it is possible to avoid air entrapment (as is the case for the experimental data reported in Figs. 2 and 4). Detailed understanding of the equilibrium morphology of the entrapped air in the encapsulated structure, i.e., whether the air is in direct contact with the core droplet in the wrapped state or the outer wrapping shell individually wraps them in a compartmentalized manner (similar to our previous demonstrations with multicomponent encapsulated droplets [35,47] within the same shell) remains a subject matter for a separate dedicated study. However, the presence of the trapped air can be confirmed from both the XZ projection of Figure 5c (i) and the side view image of Figure 5c (iv).

We repeat the same visualization procedure for double-layered encapsulated cargo. The double-layered encapsulated droplet was generated by impinging a compound droplet comprising EG as the inner core and laser oil as the outer core onto a floating interfacial layer of canola oil. The morphology of the droplet is illustrated schematically in Figure 5a (ii). Similar to the single-layered encapsulated droplet, the canola oil interfacial layer was dyed red with a red fluorescent dye. A water-soluble green dye (Bright Dyes Fluorescent FLT Yellow/Green, Kingscote Chemicals, USA) was added to EG. For imaging, two lasers (405 nm and 561 nm) were used to raster scan the droplet, and the emission signal was collected via two different detector channels. To detect the green fluorophore, we dedicated a detector channel ("Ch-Green") with a detection wavelength range of 504-617 nm, while for the red fluorophore, we used another channel ("Ch-Red," detection window 597-700 nm). Illustrative Z-slices are presented in Figure 5b (ii). Because there are two different imaging channels for the double-layered encapsulated droplet compared to only one for the single-layered droplet, we present the Z-slices for "Ch-Red", "Ch-Green" and an overlay of both the channels ("Ch-Merged") side-by-side in Figure 5b (ii). Observations similar to the single-layered droplet can also be made for the double-layered encapsulated droplet. In the case of the double-layered encapsulated droplet, $d = 1100$ μ$m$ marks the occurrence of $D_{proj}^{max}$. The enclosed inner core (dyed green), can also be visualized in the slices captured by "Ch-green." In Figure 5d, orthographic projections of the encapsulated droplet constructed from the acquired slices are presented. Similar to the orthographic projections of the single-layered droplet shown in Figure 5c, vertical cut sections along the central plane of the droplet were taken for obtaining XZ (Figure 5d (i) and YZ (Figure 5d (iii)) projections, while for XY (Figure 5d (ii)) projection, a horizontal section along $D_{proj}^{max}$ was taken. Figure 5d (iv) shows the side view of the double-layered encapsulated cargo. Air entrapment also occurs in this case and can be visualized in Figures 5d (i) and 5d (iv). The EG inner core and the outer rim (wrapping layer) of Canola oil can also be seen in both Figure 5d (i) and 5d (iv). The illustrative test cases presented in Figure 5 show how fluorescent imaging can reveal complex internal morphologies of multi-layered encapsulated droplets that would otherwise be impossible to visualize using conventional optical means.

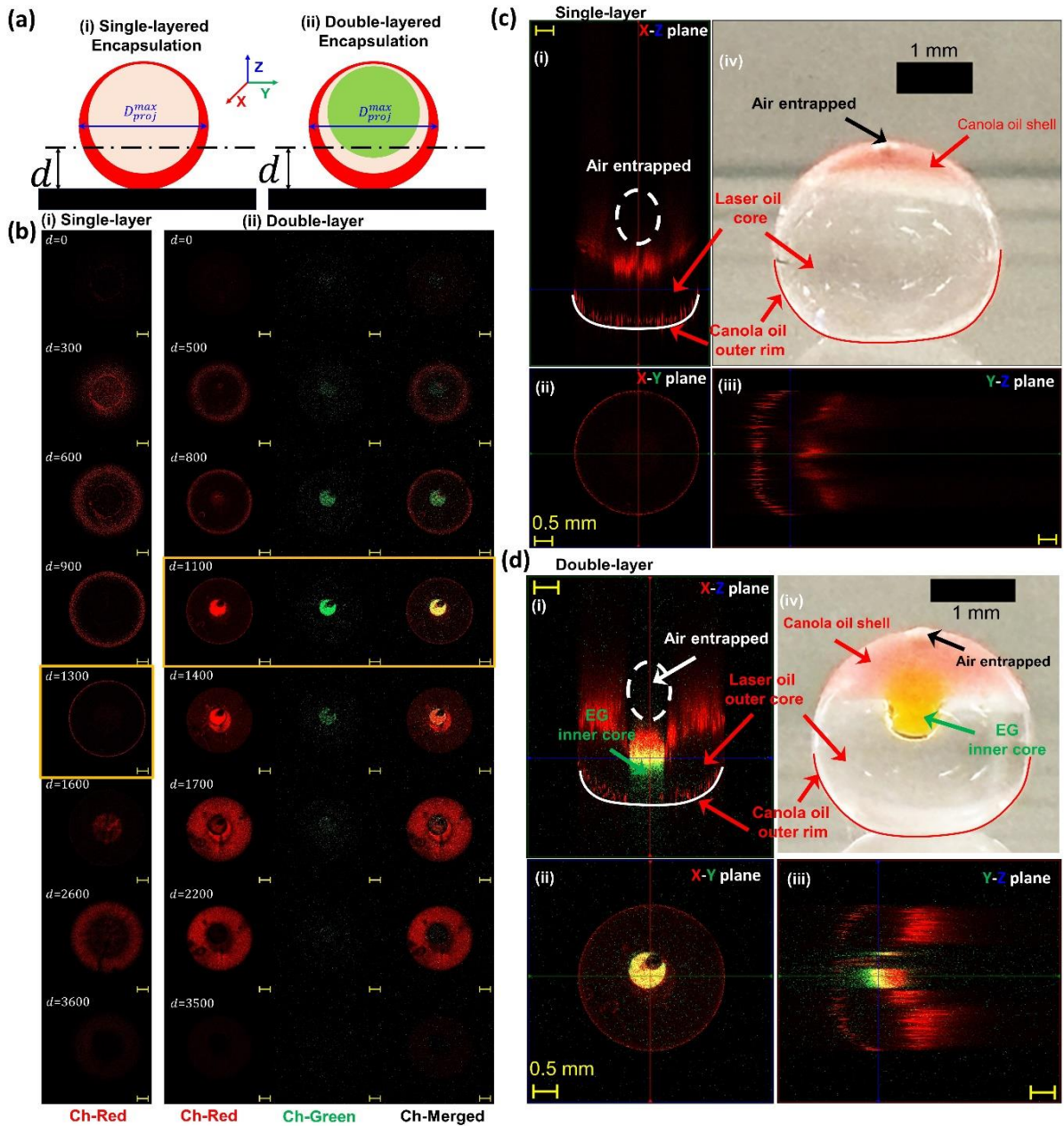

**Figure 5:** Fluorescent Visualization of the Wrapping Layer in Encapsulated Droplets: (a) Schematic of the droplet morphology for (i) Single-Layered Encapsulation, (ii) Double-Layered Encapsulation. (b) Z-stack confocal images of the (i) Single-layered encapsulated droplets and (ii) Double-layered encapsulated droplets, the scale bar in all Z-stack images represents 500 µm. (c) Orthographic projection corresponding to the Z-stack images of the single-layered encapsulated droplet shown in b (i). (i) Shows the XZ projection, while (ii) and (iii) present the XY and YZ projections, respectively, and (iv) shows the side-view optical image of the same single-layered encapsulated drop. (d) Orthographic projection corresponding to the Z-stack images of the double-layered encapsulated droplet shown in b (ii). (i) Shows the XZ projection, while (ii) and (iii) present the XY and YZ projections, respectively, and (iv) shows the side-view optical image of the same double-layered encapsulated drop. The scale bar in all Z-stack images represents 500 µm. The dark regions in the fluorescent confocal micrographs correspond to entities whose emission wavelength does not fall within the detection windows of the detector channels employed.

# 4. Conclusions

In summary, this work presents a holistic framework of liquid-liquid encapsulation to fabricate multi-layer encapsulated cargo using the combination of Y-junction and impact-driven liquid-liquid encapsulation. Compound droplets consisting of a liquid inner core and another liquid outer core are generated using the Y-junction geometry. Impingement of the compound droplet on a liquid interfacial layer (shell) floating on the host liquid bath results in wrapping the compound droplet by the liquid shell, forming the double-layered encapsulated cargo. Using time-resolved high-speed micrographs, we also discussed the underlying interfacial dynamics of the liquid-liquid interfaces during the encapsulation process. The complex interplay between interfacial (surface) tension forces, viscous forces, and momentum at the five fluids (air, interfacial layer, host liquid, outer core, and inner core) interface governs the encapsulation process. We have performed a large set of experiments with three different interfacial layer liquids to identify a non-dimensional experimental regime for encapsulation, which depicts the effect of the impact Weber number $We_i$, the viscosity ratio between the interfacial layer and the outer core $\alpha$, and the non-dimensional thickness of the interfacial layer ($\delta^*$) on the successful underwater formation of multi-layer encapsulated droplets via interfacial penetration. We observed that depending on the $We_i$, the compound droplets can either penetrate through the interfacial layer or it can get trapped at the interfacial layer and the $We_i$ that defines the boundary is identified as the critical $We_i$. We found that with the increase in viscosity ratio $\alpha$, the critical $We_i$ increases, indicating the enhancement of the viscous resistance offered by the interfacial layer during the encapsulation process. Similarly, it is found that the increase in the thickness of the interfacial layer also results in an increase in the critical $We_i$. Finally, to demonstrate the universality of the technique, we have used various interfacial layer liquids to generate the double-layered encapsulated cargo. For example, by employing an oil-based ferrofluid as the interfacial layer, we generated a magneto-responsive encapsulated droplet that can be exploited for magnet-assisted manipulation. We further extend this technique to demonstrate triple-layered encapsulation by impinging the compound droplet on two interfacial layers floating on the host water bath. Compared to the droplet microfluidics-based techniques for multi-layer encapsulation [24–30], our method caters to a much broader range of core drop volume and allows switching between drop sizes without major changes in the experimental embodiment. We show here that the present technique can overcome several functional challenges associated with original impact-driven liquid-liquid encapsulation techniques [34–37] in the context of producing multi-layer encapsulated cargo. First, introducing a carrier phase (i.e., the outer core) via Y-junction geometry eases the attainment of the necessary kinetic energy for successful interfacial penetration, thereby allowing controllable encapsulation of smaller-sized inner core droplets with a relatively lower impact height, which is otherwise challenging in original impact-driven approach [34–38]. Second, the formation of the first wrapping layer around the inner core via Y-junction could allow us to achieve multi-layered compound droplet morphologies, not achievable via the original impact-driven technique due to the requirement of a layering order of the shell-forming liquids (i.e., the interfacial layers) which is challenging to achieve because of thermodynamic unfavorability (i.e., adverse density difference or surface/interfacial tension values between shell-forming liquids). Furthermore, the carrier phase (i.e., the outer core) also protects water-soluble inner cores, resulting in lower possibilities of dissolution of the inner cargo in the host aqueous bath in the process of encapsulation and, therefore, higher success rate and repeatability in the encapsulation of aqueous cores. We also show a demonstrative example of how, with a suitable selection of fluorophores, we can use fluorescence imaging to reveal the complex internal morphology of multi-layered encapsulated cargo. The encapsulation technique presented here can further be exploited to encapsulate the target core in pH-responsive polymers and functionalize the shell material for various targeted drug delivery applications.

## Acknowledgements

U.B. and S.M. contributed equally to this work. This work was supported by the S.K.M's Discovery Grant (RGPIN-2024-03729) from the Natural Sciences and Engineering Research Council (NSERC), Canada, and the startup grant from the University of Waterloo.

## Conflict of Interest

The key liquid-liquid encapsulation technology has led to a startup SLE Enterprises B.V. Both S.M. and S.K.M have equity stakes in the startup.